\documentclass[
10pt, 
a4paper, 
oneside, 
headinclude,footinclude, 
BCOR5mm, 
]{scrartcl}

\usepackage[
nochapters, 
beramono, 
eulermath,
pdfspacing, 
dottedtoc 
]{classicthesis} 

\usepackage{hyperref}

\usepackage{arsclassica} 

\usepackage[T1]{fontenc} 

\usepackage[utf8]{inputenc} 

\usepackage{graphicx} 
\graphicspath{{Figures/}} 

\usepackage{enumitem} 

\usepackage{lipsum} 

\usepackage{subfig} 

\usepackage{amsmath,amssymb,amsthm} 

\usepackage{varioref} 

\theoremstyle{definition} 

\theoremstyle{plain} 

\theoremstyle{remark} 

\hypersetup{
colorlinks=true, breaklinks=true, bookmarks=true,bookmarksnumbered,
urlcolor=webbrown, linkcolor=RoyalBlue, citecolor=webgreen, 
pdftitle={}, 
pdfauthor={\textcopyright}, 
pdfsubject={}, 
pdfkeywords={}, 
pdfcreator={pdfLaTeX}, 
pdfproducer={LaTeX with hyperref and ClassicThesis} 
}

\title{\Large\normalfont\spacedallcaps{Comparison of the Symmetric Hyperbolic 
Thermodynamically Compatible framework with Hamiltonian mechanics of binary 
mixtures}} 

\author{Martin Sýkora\thanks{Corresponding author: 
ma.sykora@seznam.cz}\,\,\thanks{Mathematical 
Institute, Faculty of Mathematics and Physics, Charles University, Czech 
Republic},
Michal Pavelka\footnotemark[2],\,
Ilya Peshkov\thanks{Department of 	Civil, Environmental and Mechanical 
Engineering, University of Trento, Italy},\\
Piotr Minakowski\thanks{Institute of Analysis and Numerics, Otto-von-Guericke 
University 
Magdeburg, Universit\"atsplatz 2, 39106 Magdeburg, Germany},
Václav Klika\thanks{Department of Mathematics, Faculty of Nuclear Sciences and 
Physical Engineering, Czech Technical University in Prague, Prague, Czech 
Republic},
Evgeniy Romenski\thanks{Sobolev Institute of Mathematics, 4 Acad. Koptyug 
Avenue, 630090, Novosibirsk, Russia}
}

\date{} 

\usepackage{hyperref,comment}

\renewcommand{\phi}{\varphi}

\newcommand{\uu}{\mathbf{u}}
\newcommand{\JJ}{\mathbf{J}}
\newcommand{\oomega}{\boldsymbol{\omega}}
\newcommand{\ww}{\mathbf{w}}
\newcommand{\vv}{\mathbf{v}}
\newcommand{\rr}{\mathbf{r}}

\newcommand{\mm}{\mathbf{m}}
\newcommand{\qq}{\mathbf{q}}

\newcommand{\intdr}{\int\,\mathrm{d}\rr}

\begin{document}


\lehead{\mbox{\llap{\small\thepage\kern1em\color{halfgray} 
\vline}\color{halfgray}\hspace{0.5em}\rightmark\hfil}} 

\pagestyle{scrheadings} 


\maketitle 

\vspace{1.5em}
\textit{"We speak different languages, as usual," responded Woland, "but this 
  does not change the things we speak about. Well?..."}\\
\begin{flushright}\textit{--M. Bulgakov \cite{master}}\end{flushright}



\begin{abstract}
  How to properly describe continuum thermodynamics of binary mixtures where
  each constituent has its own momentum? The Symmetric Hyperbolic 
  Thermodynamically Consistent (SHTC) framework and Hamiltonian mechanics in the
  form of the General Equation for Non-Equilibrium Reversible-Irreversible 
  Coupling (GENERIC) provide two answers, which are similar but not identical, 
    and are compared in this article. They are compared both analytically and 
  numerically on several levels of 
  description, varying in the amount of detail. 
    The GENERIC equations, stemming from the Liouville equation, contain terms 
  expressing self-advection of the relative velocity by itself, which lead to a 
  vorticity-dependent diffusion matrix after a reduction. The SHTC equations, 
  on the other hand, do not contain such terms. We also show how to formulate a 
  theory of mixtures with two momenta and only one temperature that is 
  compatible with the Liouville equation and possesses the Hamiltonian 
  structure, including Jacobi identity.
\end{abstract}

\setcounter{tocdepth}{2} 

\tableofcontents 

\section{Introduction}
Homogeneous mixtures (for instance the air, petrol, aquous solutions, or plasma in fusion reactors) are ubiquitous in everyday life and industry. 
Although their behaviour can in most cases be well explained by means of equilibrium thermodynamics or fluid mechanics \cite{landau5, landau6}, there are processes that are difficult to describe theoretically. 
For instance, when a two species are mixing with high relative velocity, dynamics of the relative velocity affects the overall behaviour of the mixture (stress, heat flux, or possibly even rates of chemical reactions \cite{pavelka-ijes}). 
Our goal is to derive dynamics of binary mixtures with evolving relative velocity. A particular feature of our model is that the relative velocity is advected by itself when it has non-zero vorticity. After a reduction, this leads to a diffusion equation with a vorticity-dependent diffusion matrix. Consequently, the diffusion flux does not need to be aligned with the gradient of chemical potential, in constrast with the standard theory of mixtures \cite{dgm}.

Binary homogeneous mixtures can be described by various non-equilibrium thermodynamic 
models. These models can be classified based on the chosen state variables (fields that are equipped with their own time-evolution equations). 
In Classical Irreversible Thermodynamics (CIT), the mixture models have only 
one momentum density or velocity, typically the barycentric one \cite{dgm}, or they have none, being in the mechanical equilibrium \cite{kjelstrup}. 
In Rational Thermodynamics each constituent of the mixture can be equipped with 
its own momentum \cite{bowen-mixtures}.
In Extended Rational Thermodynamics and Extended Irreversible Thermodynamics 
the constituents can also have different temperatures 
\cite{muller-thermodynamics,muller-ruggeri,jou-eit}. 
These classical approaches take the conservation laws of mass, momentum, and 
energy for granted while seeking the undetermined fields like the stress tensor, 
heat flux, and diffusion flux via an entropic closure. There is, however, no unique 
way how to find such closure, which is why the theory of mixtures is still an 
active area of research. Alternative approaches exploit variational principles and the Lagrange$\rightarrow$Euler transformation \cite{Gavrilyuk-two}, which are the directions where we also aim.

A closure found in \cite{pavelka-ijes} leads to 
a mixture with several velocities and only one temperature, allowing for the 
diffusion to affect chemical reactions and for identification of partial 
pressure for non-ideal mixtures \cite{samohyl2012}. Especially in the 
non-isothermal case, however, there are many possibilities of finding the 
closures, and construction of the models is rather cumbersome when relying on 
the closures. In \cite{Soucek-mixtures} another theory was developed, 
leading to similar results, but facing difficulties when identifying the 
partial pressure of ideal gases in equilibrium. Yet another approach was 
developed in \cite{bothe-dreyer}, using a definition of time-reversal parity 
dependent on the choice of the physical units, making the Onsager-Casimir 
reciprocal relations dependent on the units as well. In \cite{hirschfelder} a 
theory of mixtures is obtained from the kinetic theory of rarefied gases, 
similarly as in \cite{dgm}, but equations for partial momenta of all components are not provided. 
All these theories of mixtures are based on  
entropic closures and share the property that the closure can not be 
identified uniquely. 

In this article, on the other hand, we show different ways towards theory of mixtures, relying on geometric mechanics or on the 
requirement of hyperbolicity.
We present two approaches 
to binary mixtures: (i) the 
Symmetric Hyperbolic Thermodynamically Compatible (SHTC)
equations, which is a set of first-order quasilinear hyperbolic equations 
\cite{Romensky-hyperbolic,shtc-generic,ROMENSKI2020104587}, 
and (ii) the General Equation for Non-Equilibrium Reversible-Irreversible 
Coupling (GENERIC), 
which relies on geometric mechanics, in particular Hamiltonian mechanics 
\cite{go,og,hco,pkg}.

The SHTC system can be seen as 
constructed from a Lagrangian master system of equations guaranteeing its mathematical 
properties \cite{godr,God-Siberian,GodRom-elements}. 
GENERIC, on the other hand, is constructed from four geometric building blocks: 
a Poisson bracket, an energy functional, a dissipation potential, and an 
entropy functional.
Both approaches are compatible for 
the simple continuum \cite{shtc-generic}, where the SHTC form of governing 
equations possesses the GENERIC structure. 
In the case of (binary) 
mixtures, the SHTC equations still possess the GENERIC structure. However, the 
projection from the Liouville equation (for the N-particle distribution 
function) within GENERIC leads to additional terms expressing self-advection of 
the relative velocity, which are not observed in the SHTC framework. 
We demonstrate the effects of those additional terms 
both analytically and numerically in this article.

On top of the two momenta, binary mixtures can have two entropies (or temperatures). 
By direct projection from the (Hamiltonian) Liouville equation, we obtain a Hamiltonian theory of mixtures 
where each constituent has its own density, momentum density, and entropy 
density (or temperature) \cite{pkg}. But models with only one entropy density are more 
difficult to obtain because it is not straightforward to 
fulfil the Jacobi identity, which expresses the 
self-consistency of Hamiltonian mechanics \cite{fecko}. 
Therefore, we introduce a Jacobi closure relation for the entropy difference, keeping only the overall entropy of the mixture, so that Jacobi identity is satisfied.

In Sec.\,\ref{sec2}, we recall the SHTC approach to binary mixtures. 
In Sec.\,\ref{sec3}, we first construct the GENERIC model for binary mixtures by 
projection from the Liouville equation and, subsequently, we compare it with 
the SHTC model. Moreover, a model with only one entropy is constructed via the Jacobi closure.
In Sec.\,\ref{sec4}, some difference between reduced versions of the 
SHTC and GENERIC models are discussed and compared both analytically and numerically. 

\section{SHTC framework} \label{sec2}

The SHTC equations are a class of partial differential equations of continuum 
physics with two crucial properties. Firstly, they are symmetric 
hyperbolic in the sense that they can be written, after a possible change of 
variables, in the form
\begin{equation}\label{eq.quasilin}
    A(\qq)\cdot\frac{\partial \qq}{\partial t} + B_n(\qq)\cdot \frac{\partial 
    \qq}{\partial x_n} 
    +C(\qq)\cdot\qq = 0,
\end{equation}
with $\qq$ being the vector of state variables and where the matrices $A$ and 
$B_n$ are symmetric and $A$ is positive definite 
\cite{friedrichs,Dafermos2016,Serre2007}.
One of the reasons to work with such 
restricted class of PDEs is that hyperbolic first order systems guarantee 
finite speed of propagation, unlike second order parabolic equations such as 
Navier-Stokes, and thus do not violate the concept of relativistic causality. 
Another reason is that 
\textit{symmetric} hyperbolic systems are 
locally well posed\footnote{Note that only hyperbolicity is not enough to establish the local in-time well-posedness of the Cauchy problem for a general quasilinear system \eqref{eq.quasilin} with 
non-symmetric matrices \cite{Serre2007}.}, since a unique and stable solution exists at least locally in time \cite{Dafermos2016,Serre2007}.
Moreover, the SHTC equations are compatible with thermodynamics because they satisfy the 
first and second laws of thermodynamics. 

Due to the restriction of SHTC to first order PDEs, 
classical models with higher derivatives, such as Navier-Stokes equations, 
are not within the framework directly. However, it has been shown via an asymptotic expansion in a 
relaxation parameter that solutions to the Navier-Stokes-Fourier model are approximated by the 
solutions to the SHTC equations \cite{ader-vis,romenski-viscous}.

We focus on a simplified setting close to the thermal equilibrium that can be 
modelled with just one entropy, see \cite{rbp}, and on homogeneous mixtures, 
(neglecting volume fractions). Therefore, we aim neither at
multi-temperature mixtures, like electrons and ions in low-temperature plasma, 
nor at heterogeneous mixtures, like water vapour with droplets. The following section addresses binary one-temperature homogeneous mixtures within the SHTC framework. 

\subsection{SHTC equations for binary mixtures}

Let us consider a homogeneous binary mixture composed of fluids of two 
kinds, denoted as species $1$ and $2$. A possible set of state variables 
describing 
the mixture are the pairs of densities, momentum densities, and entropy 
densities of the two species,
\begin{subequations}\label{variables}
\begin{align}
 & \rho_1, \rho_2, & \mbox{Eulerian mass densities of the two components,}\\
 &\uu_1, \uu_2, &\mbox{momentum densities of the two components,} \\
 & s_1, s_2, & \mbox{Eulerian entropy densities (per volume) of the two 
 components.}
\end{align}
\end{subequations}
Within the SHTC framework, binary mixtures are expressed in terms of the following state variables
\begin{subequations}\label{eq.var.SHTC}
        \begin{align}
        \rho_1 &= \rho_1 , & \mathrm{mass \ density \ of \ the \ first \ 
        component} , \\
        \rho&=\rho_1 + \rho_2, & \mathrm{(total) \ mass \ density}, \\
        \mm &= \uu_1 + \uu_2, &\mathrm{(total) \ momentum \ density} ,\\
        \mathbf{w}&=\frac{\uu_1}{\rho_1} - \frac{\uu_2}{\rho_2}, & 
        \mathrm{velocity \ difference},\\
        s& = s_1 + s_2. &\mbox{total entropy density.}
        \end{align}
\end{subequations}
Their SHTC time-evolution equations are
\begin{subequations}\label{eq.SHTC_full}
	\begin{align}
	\frac{\partial \rho_1}{\partial t} = &-\partial_i (\rho_1 E_{m_i}) - 
	\partial_iE_{w_i}, \label{1shtcfull} \\
	\frac{\partial \rho}{\partial t} = &-\partial_i (\rho E_{m_i}), 
	\label{2shtcfull} \\
	\frac{\partial m_i}{\partial t} = &-\rho \partial_i E_{\rho} - \rho_1 
	\partial_i E_{\rho_1} - s\partial_i E_s - m_j \partial_i E_{m_j} - \partial_j 
	(m_i E_{m_j})   \label{3shtcfull} \\ &+ E_{w_j}\partial_i w_j  - \partial_j 
	(w_i E_{w_j}), \nonumber \\
	\frac{\partial w_i}{\partial t} = &- \partial_i E_{\rho_1} -E_{m_j} \partial_j w_i 
	 - w_j \partial_i E_{m_j} - \frac{1}{\tau}E_{w^i}, 
	\label{4shtcfull}    \\ 
	\frac{\partial s}{\partial t} = &-\partial_i(s E_{m_i}) 
	+\frac{1}{\tau}\frac{1}{E_s}(E_{\ww})^2, \label{5shtcfull}
	\end{align} 
\end{subequations}
where $\tau$ is a positive relaxation parameter specifying the rate of 
dissipation, see \cite{ROMENSKI2020104587}. The dissipation is purely algebraic (involving no spatial gradients). 

In order to close the system of evolution equations, we have to specify the 
energy and the relaxation parameter. The theory so far has been independent of 
the physical system under consideration and thus general. The 
energy can be obtained by the means of statistical physics or by geometrical 
arguments based on the maximisation of entropy, see \cite{pkg}. The dissipation parameter can be obtained for instance by a dynamic reduction \cite{pkg}, by the fluctuation-dissipation theorem \cite{hco}, or, of course, it can be measured experimentally \cite{mason}.

For binary mixtures the energy is composed from the kinetic and 
the internal energy, 
\begin{align}
    E = \underbrace{\frac{\uu_1^2}{2\rho_1}+\frac{\uu_2^2}{2\rho_2}}_{=E_{\mathrm{kin}}(\rho_1, \rho, \mm, \mathbf{w})} + 
    E_{\mathrm{int}}(\rho_1, \rho, s) 
    = \frac{\mathbf{m}^2}{2\rho} + 
    \frac{\rho_1(\rho-\rho_1)}{2\rho}\mathbf{w}^2 + E_{\mathrm{int}} (\rho_1, 
    \rho,s). \label{en}
\end{align}
Therefore, the velocity is conjugate to the momentum density through 
$E_{\mathbf{m}} = \mathbf{m}/ \rho = \mathbf{v}$. 
Once the internal energy is prescribed, for instance by the Sackur-Tetrode relation 
\cite{callen} or in any other way \cite{ader-vis}, the energy becomes an 
explicit function of the state variables. Both the energy and $\tau$ will be 
chosen later in Sec. \ref{sec4} explicitly. 
In the following section we discuss the particular properties of SHTC equations like hyperbolicity.


\subsection{Properties of the SHTC Equations}
Evolution equations \eqref{eq.SHTC_full} can be written as a symmetric quasilinear system 
\eqref{eq.quasilin} of
first-order partial differential equations \cite{Romenski2007,rbp,shtc-generic}. Therefore, one can 
expect local in-time well-posedness of the Cauchy problem for \eqref{eq.SHTC_full}, provided that we have regular initial data  \cite{Serre2007} and convex energy $E$.
The symmetrization of \eqref{eq.SHTC_full} can be accomplished by introducing a generating potential 
and generating variables, as described, for example, in \cite{Romenski2009}. It should be noted 
that for the presented version of SHTC equations \eqref{eq.SHTC_full}, it is difficult to define a 
physically reasonable internal energy that is convex over a wide range of state parameters. To 
formulate the SHTC model, in which the energy is a convex function, it is necessary to extend the 
system \eqref{eq.SHTC_full} by including the equation for the evolution of the volume fraction, as 
it is done in \cite{Romenski2009}, and to define the internal energy of the mixture as the mass 
averaged internal energy of constituents' internal energies.
In this paper, however, we exclude the volume fraction from the set of state variables and consider 
simplified versions of both SHTC and GENERIC equations in order to focus on the relative velocity 
evolution equation, which is the main equation describing transport of the constituents.


Furthermore, the reversible part of the SHTC equations \eqref{eq.SHTC_full}, obtained by 
disregarding the 
relaxation terms with the prefactor $1/\tau$, is a particular realization of Hamiltonian mechanics 
\cite{shtc-generic}. 
The validity of the Jacobi identity can be shown for instance using program \cite{kroeger2010}. 

Finally, system \eqref{eq.SHTC_full} admits the total energy conservation law, see 
\cite{Romenski2007,rbp,shtc-generic}, and therefore it is compatible with the first law of 
thermodynamics. Additionally, the irreversible part of \eqref{eq.SHTC_full} (the relaxation terms) 
can be seen as the gradient of a quadratic dissipation potential, see \cite{shtc-generic}, and 
therefore the entropy production is guarantied. This implies the compatibility with the second law 
of thermodynamics.

\section{GENERIC framework} \label{sec3}

Within the GENERIC framework, summarized in monographs \cite{hco,pkg}, the time evolution of any 
state variables $\qq$ is split into reversible and irreversible parts. 
The reversible part is generated by a Poisson bracket $\{\bullet,\bullet\}$ with the Hamiltonian (or 
energy) of the system $E$, while the irreversible part is described by a 
dissipation potential $\Xi,$ and entropy $S$. 
The time evolution of any functional $A$ of the state variables is given by
\begin{equation}\label{eq.generic}
    \dot{A} = \{A, E\} + \langle A_\qq, \Xi_{\qq^*} \rangle,
\end{equation}
where subscripts stand for functional derivatives, angle brackets for the $L_2$ 
scalar product, and 
    $\qq^*=S_{\qq}$ are the conjugate variables.
The evolution 
equation for a state variable $q^i$ is obtained by plugging $A=q^i$ into Eq. \eqref{eq.generic}.

Let us now discuss the GENERIC building blocks in more detail. 
The Poisson bracket is a skew-symmetric bilinear operator, $\{F,G\}  = -\{G,F\}$,
and it satisfies the Leibniz rule and the Jacobi identity, 
\begin{align}
    &\{F,GH\}  = G\{F,H\} + \{F,G\}H\\
    &\{F,\{G,H\}\} + \{G,\{H,F\}\} + \{H,\{F,G\}\} = 0
\end{align}
for any $F(\qq)$ and $G(\qq)$. In particular, the skew-symmetry implies conservation of energy.
When the dissipation 
potential is convex with a minimum at zero in conjugate 
variables\footnote{see \cite{nonconvex} for non-convex dissipation potentials}, entropy is being produced, which guarantees compatibility with the second law of thermodynamics. 
An interesting feature of GENERIC is the fact that once the given requirements 
are met, the first and second laws of thermodynamics are automatically 
satisfied by construction, so every GENERIC model is thermodynamically 
compatible. 

The dissipation potential is usually assumed to be quadratic, which makes it compatible with the concept of dissipative brackets \cite{mor,hco}, although it does 
not have to be. 
The Poisson bracket can be obtained by a series of 
projections. For instance, Hamiltonian continuum mechanics can be obtained from the $N$-particle Liouville Poisson bracket, as 
in this article, or by other geometric techniques like variational principles \cite{Goldstein}, dynamics on Lie algebra duals \cite{arnold,ogul-tulczyjevi,ogul-tulzcyjevii}, or semidirect products
\cite{marsden-ratiu-weinstein,affro-cmat}. Once a set of state variables is 
declared, the Poisson bracket is usually derived in a unique way 
\cite{hierarchy}. When, moreover, entropy or energy density are among the set 
of state variables, only one of the pair $E$ and $S$ is necessary, since one 
can be then expressed in terms of the other, similarly as in the so called 
one-generator framework \cite{be}. The whole model for a chosen set of state 
variables only requires the knowledge of the energy of the system and of the 
dissipation potential, since these describe the particular material of the 
system. This is, for instance, the case when formulating SHTC within GENERIC 
\cite{shtc-generic}. In the next section, we show a GENERIC formulation of binary mixtures.

\subsection{Hamiltonian mechanics of mixtures by projection from the Liouville equation}
Let us briefly summarise the derivation of Hamiltonian mechanics for binary mixtures by 
projection from the Liouville equation. 
Any system composed of classical particles can be described by means of the 
$N$-particle 
distribution function, $f_N$, evolution of which is expressed by the Liouville 
equation \cite{pkg}. 
The Liouville equation is a particular realisation of the 
Lie-Poisson Hamiltonian 
mechanics, 
and is thus 
constructed from a Poisson bracket, called Liouville-Poisson bracket, and from 
an $N$-particle 
energy \cite{marsden-bbgky, hierarchy}.

When the system is composed of particles of two kinds, the $N-$particle distribution function can be projected onto
state variables of continuum mechanics \eqref{variables}. 
Such projection turns the Liouville Poisson bracket for the $N$-particle distribution 
function to 
another Poisson bracket for state variables \eqref{variables},
\begin{align}\label{eq.PB.bin}
    \{F,G\}^{\mathrm{binary}}&= \intdr \rho_1\left(\partial_i F_{\rho_1} 
    G_{u_{1i}}-\partial_i 
    G_{\rho_1} F_{u_{1i}}\right)\\
    &+\intdr u_{1i}\left(\partial_j F_{u_{1i}} G_{u_{1j}}-\partial_j G_{u_{1i}} 
    F_{u_{1j}}\right)\nonumber\\
    &+\intdr s_1\left(\partial_i F_{s_1} G_{u_{1i}}-\partial_i G_{s_1} 
    F_{u_{1i}}\right)\nonumber\\
    &+ \intdr \rho_2\left(\partial_i F_{\rho_2} G_{u_{2i}}-\partial_i 
    G_{\rho_2} 
    F_{u_{2i}}\right)\nonumber\\
    &+\intdr u_{2i}\left(\partial_j F_{u_{2i}} G_{u_{2j}}-\partial_j G_{u_{2i}} 
    F_{u_{2j}}\right)\nonumber\\
    &+\intdr s_2\left(\partial_i F_{s_2} G_{u_{2i}}-\partial_i G_{s_2} 
    F_{u_{2i}}\right)\nonumber,
\end{align}
see for instance \cite{hierarchy,pkg}.
The resulting reversible evolution equations (before adding dissipative 
dynamics) are
\begin{subequations}\label{eq.GEN.2}
\begin{align}
    \frac{\partial\rho_{a}}{\partial t} &= \partial_i (\rho_{a} E_{u_{a, i}}), 
    \label{1} \\
    \frac{\partial u_{a  i}}{\partial t} &= - \rho_{a}\partial_i E_{\rho_{a}} - 
    u_{a j} \partial_i E_{u_{a  j}} - \partial_j (u_{a i} E_{u_{a  j}}), 
    \label{2}\\
    \frac{\partial s_{a}}{\partial t} &= - \partial_i (s_a E_{u_{a 
    i}})\label{eq.sa},
    \end{align}
    for $a = 1,2$. The total energy of the system is denoted by $E$ and 
    the subscripts stand 
    for functional derivatives. If the energy is of an algebraic type, 
    $E=\int 
    e(\rho_1,\uu_1,s_1,\rho_2,\uu_2,s_2) d\rr$, where the volumetric energy density $e$ is a 
    smooth function of its 
    arguments, the functional derivatives become identical with the partial 
    derivatives of the 
    energy density.
\end{subequations}
State variables $\rho_1$, $\rho_2$, $\uu_1$, $\uu_2$, $s_1$, and $s_2$ are, however, not those in which the SHTC equations are usually expressed. The purpose of the following section is to adapt the variables closer to the SHTC framework.

\subsection{Transformation to the SHTC-like variables with two entropies}
Let us now transform the state variables so that the total density, total momentum density, and total entropy density are included,
\cite{Romensky-hyperbolic},
\begin{subequations}\label{eq.var.SHTC2}
        \begin{align}
        \rho_1 &= \rho_1 , & \mathrm{mass \ density \ of \ the \ first \ 
        component} , \\
        \rho&=\rho_1 + \rho_2, & \mathrm{(total) \ mass \ density}, \\
        \uu &= \uu_1 + \uu_2, &\mathrm{(total) \ momentum \ density} ,\\
        \mathbf{w}&=\frac{\uu_1}{\rho_1} - \frac{\uu_2}{\rho_2}, & 
        \mathrm{velocity \ difference,}\\
        s& = s_1 + s_2, &\mbox{total entropy density,}\\
        \Delta s&= \frac{s_1}{\rho_1}-\frac{s_2}{\rho_2},&\mbox{specific entropy 
        difference.}
        \end{align}
If we excluded the specific entropy difference $\Delta s$, we would have the 
same state 
variables as in SHTC.
\end{subequations}

Poisson bracket \eqref{eq.PB.bin} then transforms to
\begin{align}\label{eq.PB.new}
    \{F,G\}&= \{F,G\}^{(SHTC)} \\
    &+\intdr\frac{\rho}{\rho_1(\rho-\rho_1)}\left(\partial_i\frac{m_j}{\rho}-\partial_j\frac{m_i}{\rho}\right)F_{w_i}G_{w_j}\nonumber
     \\
    &+\intdr\frac{\partial_i w_j -\partial_j 
    w_i}{\rho_1}F_{w_i}G_{w_j}\nonumber\\
    &+\intdr\frac{\rho}{\rho_1(\rho-\rho_1)}\left(\partial_j\frac{\rho_1 
    w_i}{\rho}-\partial_i\frac{\rho_1 w_j}{\rho}\right)F_{w_i}G_{w_j}\nonumber\\
    &+\intdr \Delta s \left( \partial_i F_s G_{w_i} - \partial_i G_s 
    F_{w_i}\right)\nonumber\\
    &+\intdr \Delta s \left(G_{m_i}\partial_i F_{\Delta s}- F_{m_i}\partial_i 
    G_{\Delta s}\right)\nonumber\\
    &+\intdr \left(s_1\partial_i \frac{1}{\rho_1}-s_2\partial_i 
    \frac{1}{\rho-\rho_1}\right)\left(F_{\Delta s} G_{m_i}-G_{\Delta s} 
    F_{m_i}\right)\nonumber\\
    &+\intdr 
    \left(\frac{s_1}{\rho^2_1}-\frac{s_2}{(\rho-\rho_1)^2}\right)\left(\partial_i
     F_{\Delta s} G_{w_i}-\partial_i G_{\Delta s} F_{w_i}\right)\nonumber\\
    &+\intdr 
    \left(\frac{s_1}{\rho_1}\partial_i\frac{1}{\rho_1}-\frac{s_2}{\rho-\rho_1}\partial_i\frac{1}{\rho-\rho_1}\right)\left(F_{\Delta
     s} G_{w_i}-G_{\Delta s} F_{w_i}\right),\nonumber
\end{align}
where $s_1$ and $s_2$ are functions of $s$, $\Delta s$, $\rho$, and $\rho_1$ 
accordingly to 
transformation rules \eqref{eq.var.SHTC2}. The SHTC part of the bracket is
\begin{align}
    \{F,G\}^{(SHTC)}&=
    \intdr \rho_1 \left(\partial_i F_{\rho_1} G_{m_i}-\partial_i G_{\rho_1} 
    F_{m_i}\right)\\
    &+\intdr \rho \left(\partial_i F_{\rho} G_{m_i}-\partial_i G_{\rho} 
    F_{m_i}\right)\nonumber\\
    &+\intdr m_i \left(\partial_j F_{m_i} G_{m_j}-\partial_j G_{m_i} 
    F_{m_j}\right)\nonumber\\
    &+\intdr s \left(\partial_i F_{s} G_{m_i}-\partial_i G_{s} 
    F_{m_i}\right)\nonumber\\
    &+\intdr \left(\partial_i F_{\rho_1} G_{w_i}-\partial_i G_{\rho_1} 
    F_{w_i}\right)\nonumber\\
    &-\intdr \partial_i w_j \left( F_{w_j} G_{m_i}-G_{w_j} 
    F_{m_i}\right)\nonumber\\
    &+\intdr w_i \left(\partial_j F_{m_i} G_{w_j}-\partial_j G_{m_i} 
    F_{w_j}\right).\nonumber
\end{align}
All the terms in bracket \eqref{eq.PB.new} not involved in the part designated 
as SHTC have indeed not been contained within SHTC. As we shall see below, these terms 
imply significantly different behaviour of the two models.





Dissipation expressing mutual friction of the two species is expressed by terms
\begin{subequations}
\begin{align}
    (\partial_t w_i)_{irr} &= -\frac{1}{\tau}E_{w_i},\\
    (\partial_t s)_{irr} &= \frac{1}{\tau}\frac{1}{E_s}(E_{\ww})^2,
\end{align}
    so that the total energy is conserved by the dissipative terms and entropy 
    produced.\footnote{Note that the relaxation parameter $\tau$ is in general a symmetric 
    (by the Onsager-Casimir reciprocal relations) positive semidefinite 
    second-order twice covariant tensor even with respect to the time-reversal 
    transformation \cite{pre15}. }
\end{subequations}

The resulting system of evolution equations, including the terms expressing 
the mutual friction, reads
\begin{subequations}\label{eq.GEN-SHTC}
	\begin{align}
	\frac{\partial \rho_1}{\partial t} = &-\partial_i (\rho_1 E_{m_i}) - 
	\partial_iE_{w_i}, \label{1new} \\
	\frac{\partial \rho}{\partial t} = &-\partial_i (\rho E_{m_i}), \label{2new} 
	\\
	\frac{\partial m_i}{\partial t} = &-\rho \partial_i E_{\rho} - \rho_1 
	\partial_i E_{\rho_1} - s\partial_i E_s - m_j \partial_i E_{m_j} - \partial_j 
	(m_i E_{m_j})  \label{3new} \\ &+  E_{w_j}\partial_i w_j - \partial_j (w_i 
	E_{w_j}) - \boxed{\Delta s\partial_i E_{\Delta s} - \left( s_1 
	\partial_i \frac{1}{\rho_1} - s_2\partial_i 
	\frac{1}{\rho-\rho_1}\right)E_{\Delta s},} \nonumber \\
	\frac{\partial w_i}{\partial t} = &- \partial_i E_{\rho_1} -\partial_j w_i 
	E_{m_j} - w_j \partial_i E_{m_j} \boxed{-\Delta s \partial_i E_s }
	\label{4new}    \\ 
	& \boxed{ - \left( \frac{s_1}{\rho_1^2} - \frac{s_2}{(\rho-\rho_1)^2} 
	\right)\partial_iE_{\Delta s} - \left( 
	\frac{s_1}{\rho_1}\partial_i\frac{1}{\rho_1} - 
	\frac{s_2}{\rho-\rho_1}\partial_i\frac{1}{\rho-\rho_1} \right)E_{\Delta s} }
	\nonumber\\
	&  \boxed{ + \frac{\rho}{\rho_1 (\rho-\rho_1)} \left( \partial_i 
	\frac{m_j}{\rho} - \partial_j \frac{m_i}{\rho} \right) E_{w_j} } \nonumber \\ 
	&  \boxed{ + \frac{1}{\rho_1} (\partial_i w_j - \partial_j w_i)E_{w_j} }
	\nonumber \\ &  \boxed{ + \frac{\rho}{\rho_1 (\rho-\rho_1)} \left( 
	\partial_j \frac{\rho_1 w_i}{\rho} - \partial_i \frac{\rho_1 w_j}{\rho} 
	\right) E_{w_j}} - \frac{1}{\tau}E_{w^i}, \nonumber \\
	\frac{\partial s}{\partial t} = &-\partial_i(s E_{m_i}) 
	\boxed{-\partial_i (\Delta s 
	E_{w_i})}+\frac{1}{\tau}\frac{1}{E_s}(E_{\ww})^2, \label{5new} \\
\boxed{	\frac{\partial \Delta s}{\partial t} }= & \boxed{- 
\partial_i\left( \Delta s E_{m_i} \right) + \left( s_1 \partial_i 
\frac{1}{\rho_1} - s_2\partial_i \frac{1}{\rho-\rho_1}\right)E_{m_i} 
\label{6new}}\\ & \boxed{- \partial_i \left(\left( 
\frac{s_1}{\rho_1^2} - \frac{s_2}{(\rho-\rho_1)^2} \right)E_{w_i} \right) + 
\left( \frac{s_1}{\rho_1}\partial_i\frac{1}{\rho_1} - 
\frac{s_2}{\rho-\rho_1}\partial_i\frac{1}{\rho-\rho_1} \right)E_{w_i},} 
\nonumber
	\end{align} 
\end{subequations}
where the boxed terms are \textbf{not present in the SHTC} framework. 
In particular, we have to eliminate the $\Delta s$ variable from the equations 
in order to be compatible with the one-temperature SHTC system 
\eqref{eq.SHTC_full}.

At the moment, we can not prove the well-posedness of system \eqref{eq.GEN-SHTC} via its 
symmetrization as it can be done for the SHTC system \eqref{eq.SHTC_full}. We hope to clarify this 
question in the future. The following section contains a reduced model with only one entropy.

\subsection{Reduction to only one entropy}
The SHTC equations contain only the overall entropy $s$, not the entropy difference $\Delta s$, which is why we shall reduce the model and eliminate $\Delta s$. The elimination consists of two steps. First, only energy functionals independent of $\Delta s$ are considered, and, secondly, the entropy difference is expressed in terms of the remaining state variables, $\Delta s(\rho_1,\rho,s,\mm,\ww)$ while keeping Jacobi identity valid. The second step is called a Jacobi closure \cite{miroslav-grad}. The whole bracket \eqref{eq.PB.new} of course satisfies the Jacobi identity because it is just a transformation of the Poisson bracket \eqref{eq.PB.bin}, and if we drop the lines with $s$ and $\Delta s$, it still satisfies Jacobi identity\footnote{Jacobi identity can be checked using program \cite{kroeger2010}.}. Such bracket with no entropies is suitable for isothermal mixtures, as shown in \cite{elafif2002}. Here, however, we aim at non-isothermal mixtures, keeping $s$ among the state variables.

Since $\Delta s$ is no longer a state variable, the total energy does not explicitly depend on it, so we have
\begin{equation}\label{eq.Eds}
E_{\Delta s}=0, 
\end{equation}
which means that $E_{\Delta s} = 
\frac{\rho_1 - \rho_2}{\rho}\left( E_{s_1} - E_{s_2} \right)=0$ and consequently 
$T_1 = E_{s_1} = E_{s_2} = T_2$, which means that the temperatures of the two constituents 
are equal. Therefore the restriction to only the overall entropy is possible 
only if the two constituents of the mixture have the same temperature. This is not satisfied only in rather extreme cases of homogeneous mixtures like electrons and ions in cold plasma \cite{chen}.

The last four lines of \eqref{eq.PB.new}, which generate equation \eqref{6new}, 
disappear, but the 
fifth line of \eqref{eq.PB.new} still contains $\Delta s$, which translates to 
the boxed terms in equation \eqref{5new} and in the first line of \eqref{4new}. 
Using a program \cite{kroeger2010}, we now check whether there exists an 
admissible formula for $\Delta s$ such that the remaining terms of 
\eqref{eq.PB.new} (without the last four lines) fulfil Jacobi identity. 
We first assume a general dependence $\Delta s(\rho,\rho_1,\mm,\ww,s)$ and get 
a necessary condition that it does not depend on $\mm$ to satisfy the Jacobi 
identity. 
Then we consider $\Delta s$ independent of $\mm$ which yields that $\Delta s$ must be 
independent of $\ww$ as well. Choosing such entropy difference, 
we get that $\Delta s$ must be $0$-homogeneous, that is $\Delta s 
= \Delta s \left(\phi, \psi \right)$, where $\phi= \rho/s$ and $\psi= 
\rho_1/s$. The next iteration of this procedure gives the condition
\begin{equation}
\partial_{\psi}\Delta s - \Delta s \left( \phi\partial_{\phi} \Delta s + 
\psi \partial_{\psi} \Delta s \right) = 0, \label{partial}
\end{equation}
particular solution of which is
\begin{subequations}\label{eq.ds}
\begin{equation} \label{cond}
    \Delta s = \frac{\phi_0-\phi}{\phi_0\psi- \phi\psi_0}
\end{equation}
for arbitrary parameters $\phi_0, \psi_0$. Note, however, that this class does 
not contain all solutions, for instance the constant solution. Finally, using 
again the program 
\cite{kroeger2010}, it becomes necessary that
\begin{equation}\label{cond2}
    \psi_0=0 \quad\mbox{or}\quad \psi_0=\phi_0
\end{equation}
\end{subequations}
in order that the Jacobi identity be fulfilled. Even though we have not studied 
all possible solutions of \eqref{partial}, we have found a family of reductions 
satisfying Jacobi identity. 
The constant solution, however, does belong to that family.\footnote{Another possibility to find $\Delta s$ unambiguously is to focus directly on Eq. \eqref{eq.Eds}. Using results from \cite{pkg} for a binary mixture of two ideal gases with the same atomic masses, we obtain that $\Delta s$ should be equal to $\frac{k_B}{m} \ln\left(\frac{\rho}{\rho_1}-1\right)$, which, however, does not satisfy the Jacobi identity. }

To sum up, the above choice of $\Delta s$  (condition \eqref{cond})  leads to a simplified Poisson bracket which fulfils 
the Jacobi identity. In terms of equations, system \eqref{eq.GEN-SHTC} then simplifies to
\begin{subequations}\label{eq.GEN-SHTC-one-entropy}
	\begin{align}
	\frac{\partial \rho_1}{\partial t} = &-\partial_i (\rho_1 E_{m_i}) - 
	\partial_iE_{w_i}, \label{1one} \\
	\frac{\partial \rho}{\partial t} = &-\partial_i (\rho E_{m_i}), \label{2one} 
	\\
	\frac{\partial m_i}{\partial t} = &-\rho \partial_i E_{\rho} - \rho_1 
	\partial_i E_{\rho_1} - s\partial_i E_s - m_j \partial_i E_{m_j} - \partial_j 
	(m_i E_{m_j})  \label{3one} \\ &+ E_{w_j}\partial_i w_j - \partial_j (w_i 
	E_{w_j}) , \nonumber \\
	\frac{\partial w_i}{\partial t} = &- \partial_i E_{\rho_1} -E_{m_j} \partial_j w_i 
	 - w_j \partial_i E_{m_j} \boxed{-\Delta s \partial_i E_s }
	\label{4one}    \\ 
		&  \boxed{ + \frac{\rho}{\rho_1 (\rho-\rho_1)} \left( \partial_i 
		\frac{m_j}{\rho} - \partial_j \frac{m_i}{\rho} \right) E_{w_j} }\nonumber \\ 
	&  \boxed{+ \frac{1}{\rho_1} (\partial_i w_j - \partial_j w_i)E_{w_j} }
	\nonumber \\ &  \boxed{+ \frac{\rho}{\rho_1 (\rho-\rho_1)} \left( 
	\partial_j \frac{\rho_1 w_i}{\rho} - \partial_i \frac{\rho_1 w_j}{\rho} 
	\right) E_{w_j}} - \frac{1}{\tau}E_{w^i}, \nonumber \\
	\frac{\partial s}{\partial t} = &-\partial_i(s E_{m_i}) 
	\boxed{-\partial_i (\Delta s 
	E_{w_i})}+\frac{1}{\tau}\frac{1}{E_s}(E_{\ww})^2, \label{5one}
    \end{align}
\end{subequations}
where $\Delta s$ is a function of $\rho/s$ and $\rho_1/s$.
This is a two-fluid model with one temperature derived from the Liouville 
equation only through projections and a Jacobi closure. 
The total mass, momentum, and the total energy are conserved by 
equations \eqref{eq.GEN-SHTC} and \eqref{eq.GEN-SHTC-one-entropy}.
This model is Hamiltonian and fulfils the Jacobi identity.


There is also another way to obtain a Hamiltonian model with only the 
overall entropy. 
If we first drop the second, third and fourth lines of 
\eqref{eq.PB.new}, which represent the difference between the SHTC evolution 
and the evolution stemming from the Liouville equation, then, following the 
above procedure based on the program \cite{kroeger2010}, we get that the Jacobi 
identity is satisfied as long as $\Delta s$ is constant.
In other words, if we drop the three 
boxed terms not containing $\Delta s$, we can leave the others and reduce the 
model to single entropy $s$ as long as $E_{\Delta s}=0$ and $\Delta s$ is 
constant. If actually $\Delta s = 0$, we obtain the SHTC equations as a special 
case. Thus the SHTC model is Hamiltonian, and stays Hamiltonian even after  
adding the boxed terms containing constant $\Delta s$. However, it is derived 
from the Liouville equation not just by projections, but also by the further 
neglection of the boxed terms. The Liouville equation leads to the SHTC 
equations extended by extra terms that express self-advection of the relative 
velocity field $\ww$ and extra advection of total entropy by that field. 
This difference is demonstrated in the following section while noting that $\Delta s=0$ requires identical specific partial entropies rather than equal temperatures.

\section{Reduction and comparison of SHTC and GENERIC}\label{sec4}
In order to describe the difference between the SHTC and GENERIC models for binary mixtures, we further reduce the models so that the effects of the 
self-advection of $\ww$ are highlighted. Systems 
\eqref{eq.SHTC_full} and \eqref{eq.GEN-SHTC-one-entropy} are reduced to 
simpler systems of equations containing less detail. Assuming that the dynamics 
of the $\ww$ field is fast (relaxation parameter $\tau$ 
sufficiently small), 
solutions to the full systems of equations should be close to solutions to  
reduced systems without $\ww$ as a state variable. One way to carry out that 
reduction is asymptotic analysis \cite{ader-vis}. Here we follow 
another way, called the Dynamic Maximum Entropy reduction (DynMaxEnt), which is 
based on the geometric character of the evolution equations.

\subsection{DynMaxEnt reduction of GENERIC}
DynMaxEnt is motivated by contact geometry, and it is typically compatible with
asymptotic analysis \cite{arnoldbook,pkg,dynmaxent}.
Let us denote the energetic conjugate variables as $\rho_1^{\dagger}=E_{\rho_1}$, 
$\rho^{\dagger}=E_{\rho}$, $\mathbf{m}^{\dagger}=E_{\mathbf{m}}$, 
$\mathbf{w}^{\dagger}=E_{\mathbf{w}}$, and
$s^{\dagger}=E_{s}$.
The DynMaxEnt reduction of $\mathbf{w}$ is carried out in these steps: 
(i) Set 
$\ww$ to its maximum entropy (MaxEnt) value \cite{callen} (here $\mathbf{w}=0$), which corresponds to the minimum of energy \eqref{en}, 
(ii) calculate 
$\mathbf{w}^{\dagger}$ from the evolution equation of $\ww$, 
and (iii) plug the 
calculated $\ww$ and $\ww^\dagger$ into the remaining equations. 

From Eq.~\eqref{4one} we get that 
\begin{equation}
    \tau  \Gamma  = - (\mathbb{I} - \alpha \Omega)\mathbf{w}^{\dagger},
\end{equation}
where 
\begin{equation}\label{eq.alpha}
\Gamma = \nabla \rho_1^{\dagger} + \Delta s \nabla s^{\dagger},
\quad
\Omega_{i,j} = \partial_i m_j^{\dagger} - \partial_j m_i^{\dagger},
\quad\text{and}\quad
\alpha = \frac{\tau \rho}{\rho_1(\rho-\rho_1)}.
\end{equation}
Denoting $\Theta = (\mathbb{I} - \alpha \Omega)$, we obtain a relation for 
$\ww^\dagger$,
\begin{align}
   \mathbf{w}^{\dagger} = - \tau \Theta^{-1} \Gamma, \label{maxent}
\end{align}
that is to be substituted into the remaining equations (for $\rho$, $\rho_1$, 
and $\mm$).
By introducing quantities 
\begin{equation}
    a = -\alpha (\partial_1m_2^{\dagger} - \partial_2m_1^{\dagger}),\,
    b = -\alpha (\partial_1m_3^{\dagger} - \partial_3m_1^{\dagger}), 
    \,\text{and}\,
    c = -\alpha (\partial_2m_3^{\dagger} - \partial_3m_2^{\dagger}),
\end{equation}
which are proportional to the overall vorticity, matrix $\Theta$ can be rewritten as
\begin{equation}
    \Theta = 
\begin{pmatrix}
1 & a & b  \\
-a & 1 & c \\
-b & -c & 1
\end{pmatrix}.
\end{equation}
Moreover,
\begin{align*}
    \det(\Theta) &= 1+a^2+b^2+c^2 > 0, \ \ \ \Theta > 0, \text{ and}\\
    \Theta^{-1} &= \frac{1}{\det(\Theta)}
\begin{pmatrix}
1 + c^2 & -a-bc & -b+ac  \\
a-bc & 1 + b^2 & -c-ab \\
b+ac & c-ab & 1 + a^2
\end{pmatrix} \approx
    \begin{pmatrix}
1 & -a & -b  \\
a & 1 & -c \\
b & c & 1
    \end{pmatrix} \stackrel{\text{def}}{=} \overline{\Theta^{-1}},
\end{align*}
which is an approximate inverse of matrix $\Theta$ in the regime of low overall 
vorticity. 
Matrices $\Theta, \Theta^{-1}, \overline{\Theta^{-1}}$ are positive definite 
because their symmetric parts are positive definite although
the eigenvalues of $ \Theta $, $ \{1, 1\pm\sqrt{-a^2-b^2-c^2} \} $,  are not 
all real (bearing in mind that the matrices are not symmetric).

System of equations \eqref{eq.GEN-SHTC-one-entropy} is then reduced to 
\begin{subequations}\label{eq.red.0}
\begin{align}
    \frac{\partial \rho_1}{\partial t} =  &-\partial_i (\rho_1 m_i^{\dagger}) + 
    \nabla\cdot \underbrace{ \left( \tau \Theta^{-1} \Gamma \right)}_{=-\JJ}, 
    \label{hust.0}\\
    \frac{\partial \rho}{\partial t} = &-\partial_i (\rho m_i^{\dagger}) 
    ,\label{hmot.0} \\
            \frac{\partial m_i}{\partial t} = &-\rho \partial_i \rho^{\dagger} 
            - \rho_1 \partial_i \rho_1^{\dagger} - s\partial_i s^{\dagger} - 
            m_j \partial_i m_j^{\dagger} - \partial_j (m_i m_j^{\dagger}) 
            \label{hyb.0}  \\
    \frac{\partial s}{\partial t} = &-\partial_i(s m_i^{\dagger}) + \nabla\cdot 
    \left(\Delta s \tau \Theta^{-1} \Gamma \right) + \frac{\tau}{s^{\dagger}} 
    \left(\Theta^{-1} \Gamma \right)^{2},  \label{ent.0}
\end{align}
which no longer contains $\ww$ as a state variable.
\end{subequations}

Let us now discuss the properties of this reduced system of equations. 
Firstly,  entropy is produced because
the right-hand side of \eqref{ent.0} consists of divergence terms and positive 
terms. Secondly, equation \eqref{hust.0} can be interpreted as advection of $\rho_1$  
by the velocity and as diffusion of $\rho_1$ with a vorticity-dependent 
non-symmetric diffusion matrix.
When using the approximate inverse $\overline{\Theta^{-1}}$ and neglecting the
entropies (the isothermal case), the flux $\JJ$ becomes
$$\JJ\approx -\tau \nabla \rho_1^\dagger +\tau \alpha \oomega \times \nabla 
\rho^\dagger_1,$$
where $\oomega=\nabla\times\vv$ is the overall vorticity. The 
flux then consists of two terms, one proportional to the gradient of chemical 
potential and the other perpendicular to that gradient and the 
vorticity. We shall study the effects of the latter term later in this section.

\subsection{DynMaxEnt reduction of SHTC}

Similarly as in the case of GENERIC, DynMaxEnt reduces the SHTC equations \eqref{eq.SHTC_full} to
\begin{subequations}\label{eq.red.2}
	\begin{align}
	\frac{\partial \rho_1}{\partial t} =  &-\partial_i (\rho_1 m_i^{\dagger}) + 
	\nabla\cdot \underbrace{\left( \tau  \nabla \rho_1^{\dagger} 
	\right)}_{=-\tilde{\JJ}}, \label{hust.2}\\
	\frac{\partial \rho}{\partial t} = &-\partial_i (\rho m_i^{\dagger}) 
	,\label{hmot.2} \\
	\frac{\partial m_i}{\partial t} = &-\rho \partial_i \rho^{\dagger} - \rho_1 
	\partial_i \rho_1^{\dagger} - s\partial_i s^{\dagger} - m_j \partial_i 
	m_j^{\dagger} - \partial_j (m_i m_j^{\dagger}) \label{hyb.2},  \\
	\frac{\partial s}{\partial t} = &-\partial_i(s m_i^{\dagger}) + 
	\frac{\tau}{s^{\dagger}} \left( \nabla \rho_1^{\dagger} \right)^{2}. 
	\label{ent.2}
	\end{align}
\end{subequations}
This model has similar features as model \eqref{eq.red.0}.


The main difference between GENERIC and SHTC for binary mixtures is 
the self-advection of $\ww$. In the following section, we confine ourselves to 
the isothermal case to highlight this difference.



\subsection{The isothermal regime in two-dimensions}

The reduction of both the SHTC and GENERIC models (\eqref{eq.red.2}, 
\eqref{eq.red.0}) to isothermal case (dropping all terms with the entropy) leads to
\begin{subequations}\label{eq.red}
\begin{align}
    \frac{\partial \rho_1}{\partial t} =  &-\partial_i (\rho_1 m_i^{\dagger}) + 
    \nabla\cdot \left( \tau \Theta^{-1}  \nabla \rho_1^{\dagger} \right), 
    \label{hust}\\
    \frac{\partial \rho}{\partial t} = &-\partial_i (\rho m_i^{\dagger}) 
    ,\label{hmot} \\
            \frac{\partial m_i}{\partial t} = &-\rho \partial_i \rho^{\dagger} 
            - \rho_1 \partial_i \rho_1^{\dagger} - m_j \partial_i m_j^{\dagger} 
            - \partial_j (m_i m_j^{\dagger}) \label{hyb} ,
\end{align}
\end{subequations}
where the difference between the two models is encoded into the ``diffusion 
matrix'' $\Theta^{-1}$.
Next, we linearise the ``diffusion matrix'' $\Theta^{-1}$ to $\overline{\Theta^{-1}}$ and restrict ourselves
to two dimensions, which gives
\begin{align}
    \Theta^{-1}\approx \begin{pmatrix}
1 & -a   \\
a & 1 \end{pmatrix} 
\end{align}
with
\begin{align}
    a &= 0, &\text{for SHTC,} \\
       a &= - \alpha 
    \left( \partial_1\left( \frac{m_2}{\rho}\right) - \partial_2\left( 
    \frac{m_1}{\rho}\right) \right) = -\alpha 
    \mathrm{curl}\left(\frac{\mathbf{m}}{\rho}\right) 
    &\text{for GENERIC.}\label{generic_a}
\end{align} 

Let us now interpret the quantities $a$ and $\alpha$.
From the definition \eqref{eq.alpha} it follows that $ \alpha \sim  
\frac{\tau}{\rho} $, the units of $\alpha$ are thus seconds, and $\alpha$ 
represents the microscopic time scale due to the relative velocity $\ww$, 
$t_{\text{micro}}$. The units of the vorticity $\Omega$ are $s^{-1}$ and 
thus the vorticity represents the macroscopic time scale related to the overall 
vortices, $t_{\text{macro}}$. Quantity $a$ represents the ratio of the 
microscopic relative velocity time scale and the macroscopic vorticity time 
scale, $a \sim t_{\text{micro}}/t_{\text{macro}}$.


In order to compare both approaches quantitatively, we have to provide an energy functional and a formula for the relaxation parameter.
In the isothermal case the energy can be substituted with the free energy, see \cite{pkg}. 
Therefore, we choose a simplified free energy of two ideal gasses 
\begin{equation}
    E = \frac{\mathbf{m}^2}{2\rho} + \frac{\kappa}{\alpha} \left( \rho_1 \log 
    (\rho_1) - (\rho_1 - 
    \rho) \log(\rho-\rho_1)\right),
\end{equation}
where $\kappa$ and $\alpha$ are constants.
Furthermore, the choice of $\alpha$ also defines the  relaxation parameter 
$\tau$ through relation \eqref{eq.alpha}.

We now write \eqref{eq.red} in a compact divergence form as
\begin{subequations}\label{adult}
\begin{align}
    \frac{\partial \rho_1}{\partial t} &=  - \mathrm{div} \left(\rho_1 
    \frac{\mathbf{m}}{\rho} - 
     \tau \Theta^{-1}\nabla E_{\rho_1}\right) , \\
    \frac{\partial \rho}{\partial t} &= -\mathrm{div} (\mathbf{m}), \\
     \frac{\partial m_i}{\partial t} &= -   \mathrm{div}\left(\frac{m_i 
     \mathbf{m}}{\rho} + pe_i \right)  , i=1,2,
\end{align}
\end{subequations}
where the pressure $p$ stands for
\begin{equation}
    p = -e + \rho \frac{\partial e}{\partial \rho} + \rho_1 \frac{\partial e}{\partial \rho_1} + \mathbf{m}\cdot \frac{\partial e}{\partial \mathbf{m}}.
\end{equation}
Even more explictly, we obtain that
\begin{subequations}\label{adult2}
\begin{align}
    \frac{\partial \rho_1}{\partial t} &=  - \mathrm{div} \left(\rho_1 
    \frac{\mathbf{m}}{\rho}\right) + \mathrm{div} \left( 
     \tau \Theta^{-1}\nabla 
     \left(\frac{\kappa}{\alpha}(\log(\rho_1)-\log(\rho-\rho_1)) 
     \right)\right),\label{a2_1}\\
    \frac{\partial \rho}{\partial t} &= -\mathrm{div} (\mathbf{m}), \\
     \frac{\partial m_i}{\partial t} &= -   \mathrm{div}\left(\frac{m_i 
     \mathbf{m}}{\rho} + \frac{\kappa}{\alpha}\rho \right)  , i=1,2.
\end{align}
\end{subequations}
The pressure can be rewritten as 
$p= \frac{\kappa}{\alpha}\rho$. Evaluating gradients in the right-most term 
of \eqref{a2_1}, gives
\begin{align}
    \mathrm{div} \left( 
     \tau \Theta^{-1}\nabla
      \left(\frac{\kappa}{\alpha}(\log(\rho_1)-\log(\rho-\rho_1)) 
      \right)\right) &= \mathrm{div} \left( \tau \Theta^{-1}   \nabla\rho_1 - 
      \tau \Theta^{-1} \frac{\rho_1}{\rho}  \nabla\rho \right) \nonumber\\ 
      &= 
      \mathrm{div} \left(\kappa 
      \Theta^{-1}   \rho  \nabla \left(\frac{\rho_1}{\rho} \right)\right).
\end{align}
The system of equations then simplifies to 
\begin{subequations}\label{adult4}
\begin{align}
    \frac{\partial \rho_1}{\partial t} &=  - \mathrm{div} \left(\rho_1 
    \frac{\mathbf{m}}{\rho} - 
    \kappa \Theta^{-1}   \rho  \nabla \left(\frac{\rho_1}{\rho} \right)\right), 
    \label{rhoj4}\\
    \frac{\partial \rho}{\partial t} &= -\mathrm{div} (\mathbf{m}), 
    \label{rho4}\\
     \frac{\partial m_i}{\partial t} &= -   \mathrm{div}\left(\frac{m_i 
     \mathbf{m}}{\rho} + \frac{\kappa}{\alpha}\rho \right)  , \ i=1,2, 
     \label{u4}
\end{align}
which we solve both analytically and numerically.
\end{subequations}

\subsection{Analytical comparison}\label{sec:reduced_model}
In this section, we illustrate the difference between SHTC and GENERIC 
formulations by analytical means, solving the simplified system 
\eqref{adult4}.  
Equation 
\eqref{rhoj4} gets within the SHTC model the following form,
\begin{align} \label{shtcrhoj}
\frac{\partial \rho_1}{\partial t} &=  - \mathrm{div} \left(\rho_1 
\frac{\mathbf{m}}{\rho}\right) + 
\kappa \Delta \rho_1  - \kappa \frac{\rho_1}{\rho} \Delta \rho - \kappa \nabla 
\left( 
\frac{\rho_1}{\rho} \right) \cdot \nabla \rho.
\end{align}
On the other hand, within the GENERIC model, we obtain
\begin{align} \label{genrhoj}
\frac{\partial \rho_1}{\partial t} =  &- \mathrm{div} \left(\rho_1 
\frac{\mathbf{m}}{\rho}\right) + 
\kappa \Delta \rho_1  - \kappa \frac{\rho_1}{\rho} \Delta \rho - \kappa \nabla 
\left( 
\frac{\rho_1}{\rho} \right) \cdot \nabla \rho \nonumber \\
&+ \kappa \left( \frac{\partial a}{\partial y}\frac{\partial \rho_1}{\partial 
x} - \frac{\partial 
a}{\partial x} \frac{\partial \rho_1}{\partial y}\right) - \kappa 
\frac{\rho_1}{\rho}\left( 
\frac{\partial a}{\partial y}\frac{\partial \rho}{\partial x} - \frac{\partial 
a}{\partial x} 
\frac{\partial \rho}{\partial y}\right) \\
&- \kappa \nabla \left( \frac{\rho_1}{\rho} \right) \cdot \left( -a 
\frac{\partial \rho}{\partial y}, a \frac{\partial \rho}{\partial x} 
\right)^{\top} .  \nonumber
\end{align}
Realising that the equations \eqref{rho4} and \eqref{u4} are not coupled with 
\eqref{rhoj4}, we first search for stationary $\mm$ and 
$\rho$, solving Eqs. \eqref{rho4} and \eqref{u4}, and then for a non-stationary $\rho_1$,
solving Eq. \eqref{rhoj4}. 
The following functions indeed represent stationary solutions,
\begin{align} \label{init}
    \rho &= 1, \\
    \mathbf{v} &= (v_1,v_2) = (10 \sin(y),0),
\end{align}
where $\vv$ is the velocity field defined by $\mm=\rho \vv$. Plugging these solutions into Eqs.
\eqref{shtcrhoj} and \eqref{genrhoj}, we get
\begin{align}
    \frac{\partial \rho_1}{\partial t} &=  - \mathrm{div} \left(\rho_1 
    \frac{\mathbf{m}}{\rho}\right) + \kappa \Delta \rho_1 
    &\text{for SHTC}, \\
    \frac{\partial \rho_1}{\partial t} &=  - \mathrm{div} \left(\rho_1 
    \frac{\mathbf{m}}{\rho}\right) + \kappa \Delta \rho_1 + \kappa \left( 
    \frac{\partial a}{\partial y}\frac{\partial \rho_1}{\partial x} - 
    \frac{\partial a}{\partial x} \frac{\partial \rho_1}{\partial y}\right)&\text{for GENERIC}.   
    \label{rhoj4new}
\end{align}

We now recall the vorticity diffusion ratio $a$, that for GENERIC takes the form
\begin{equation}
a = -\alpha (\partial_x v_2 - \partial_y v_1) = \alpha \partial_y v_1.
\end{equation}
As a result, $\partial_x a = 0$ and $\partial_y a = \alpha \partial _{yy} v_1 = 
- 10 \alpha \sin(y) = - \alpha v_1$.
Plugging this into Eq. \eqref{rhoj4new} yields
\begin{align}
    \frac{\partial \rho_1}{\partial t} &= - \mathrm{div} \left(\rho_1 
    \frac{\mathbf{m}}{\rho}\right) + \kappa \Delta \rho_1  - \kappa\alpha v_1 
    \frac{\partial \rho_1}{\partial x}.
\end{align}
Since $v_1$ does not depend on $x$ and $v_2=0$, we get
\begin{align}
    \frac{\partial \rho_1}{\partial t} &= - (1 + \kappa \alpha) \mathrm{div} 
    \left(\rho_1 \frac{\mathbf{m}}{\rho}\right) + \kappa \Delta \rho_1 . 
    \label{rhojfinal}
\end{align}
In summary, both SHTC and GENERIC models contain advection and 
diffusion of $\rho_1$. The advection in GENERIC, however, has a coefficient $1+\kappa \alpha$ while 
in SHTC the coefficient equals $1$, even though the velocity field is identical in both 
cases. This is an effect of the 
vorticity-dependent part of the diffusion matrix in the reduced GENERIC model. 
Let us now examine these effects in more detail numerically.

\subsection{Numerical comparison}

\begin{figure}[ht]
  \begin{center}
    \begin{tabular}{ccc} 
      \includegraphics[trim = 10 10 10 
      10,clip,width=0.3\linewidth]{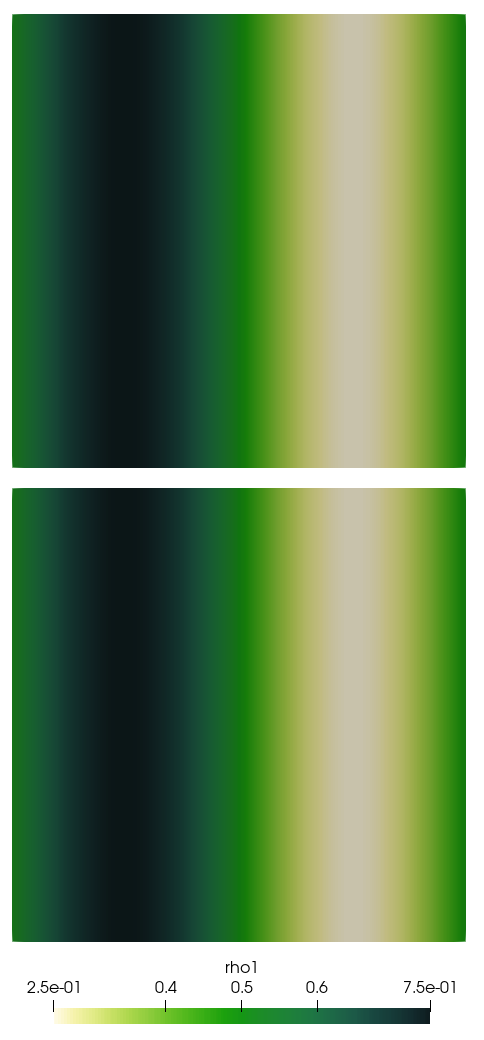}
      &
      \includegraphics[trim = 10 10 10 
      10,clip,width=0.3\linewidth]{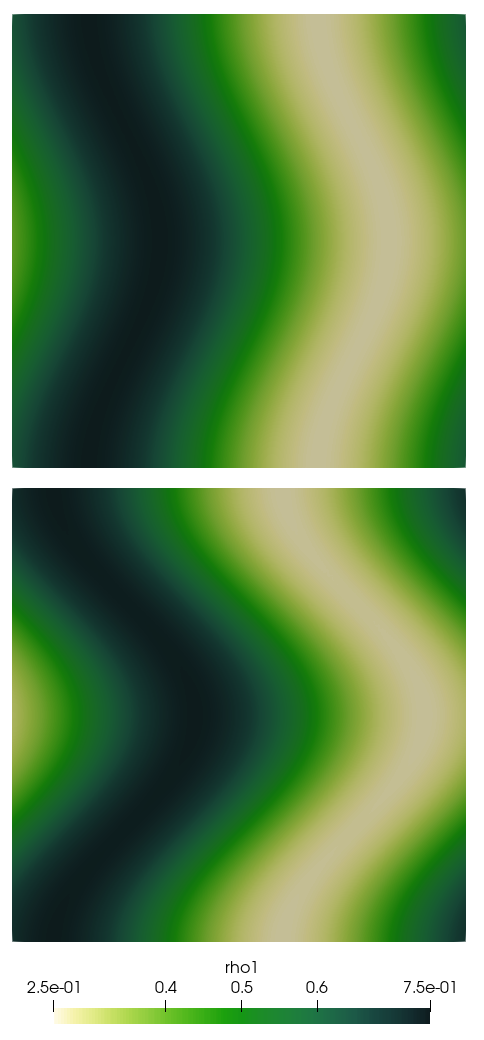} &
      \includegraphics[trim = 10 10 10 
      10,clip,width=0.3\linewidth]{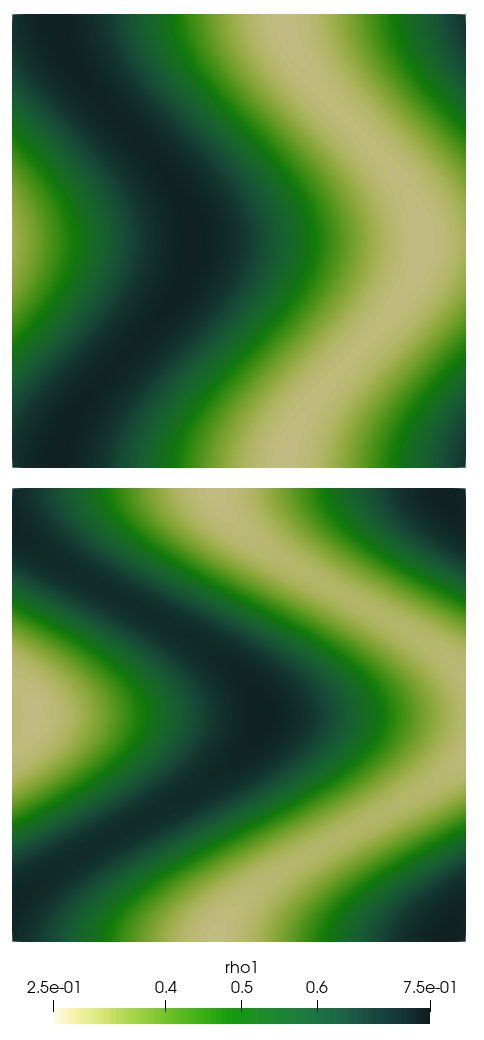}\\
      \includegraphics[width=0.3\linewidth]{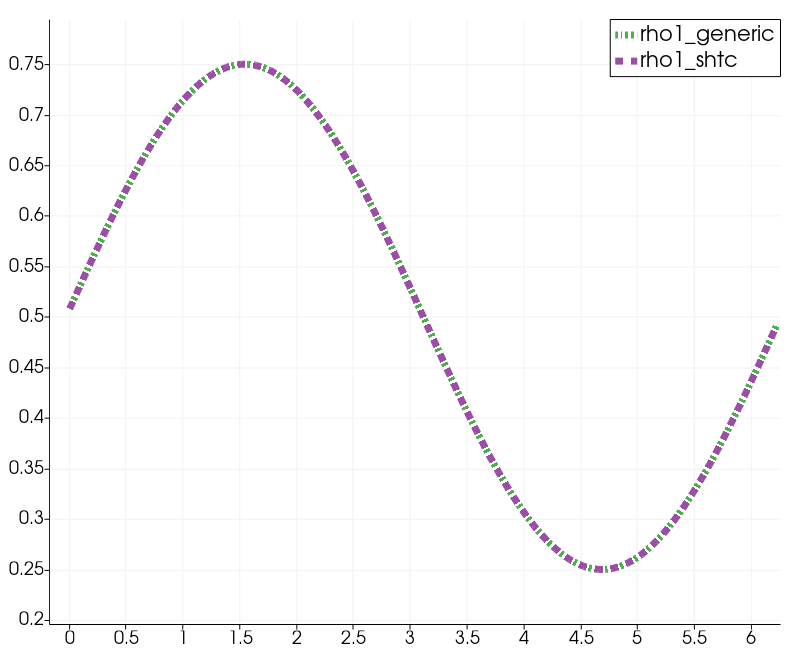}
      &
      \includegraphics[width=0.3\linewidth]{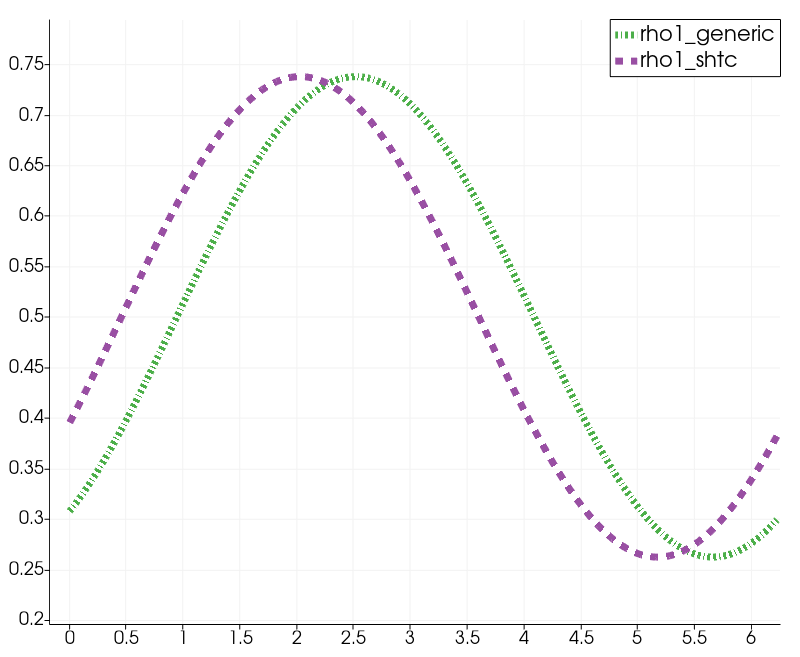} &
      \includegraphics[width=0.3\linewidth]{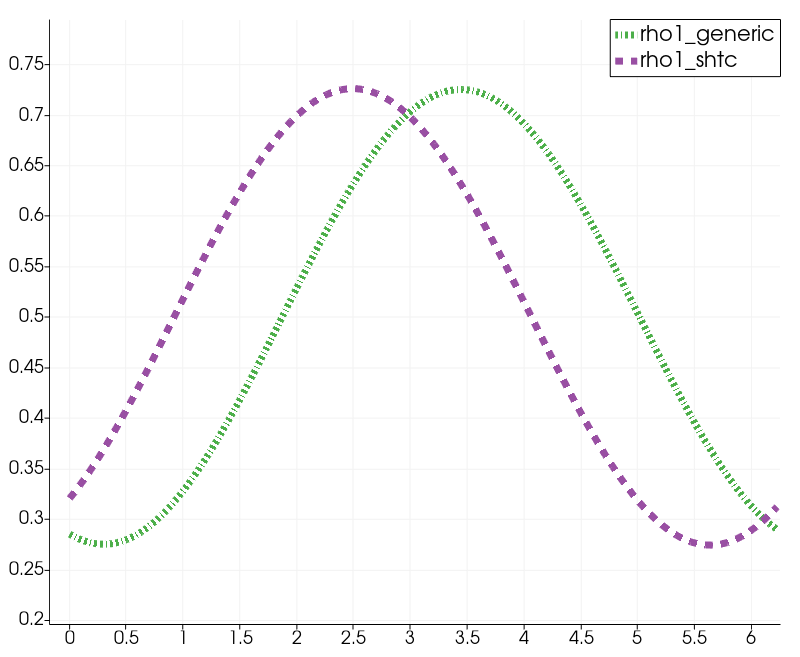}
    \end{tabular}
    \caption{Steady laminar flow of a binary mixture with periodic boundary 
      conditions. The density of the first component is diffused and advected 
      by 
      the flow, for three times (from left to right $t = 
      0,T_{\mathrm{final}}/2,T_{\mathrm{final}}$) under the SHTC 
      model (top), and the GENERIC model (middle). The bottom line 
      shows one-dimensional density profiles of the first component along 
      x-axis at the middle of the y-axis.}
    \label{fig1}
  \end{center}
\end{figure}

We now solve the reduced system \eqref{adult4} numerically in order to illustrate 
the differences between the reduced SHTC and GENERIC systems.
The numerical implementation is realised in the software library PyPDE 
\cite{Jackson-plastic}, using ADER-WENO \cite{toro,ader-vis} method for 
solution to equations in the form:
\begin{equation}
\frac{\partial \mathbf{q}}{\partial t}+\nabla \mathbf{F}(\mathbf{q}, \nabla 
\mathbf{q})+B(\mathbf{q}) \cdot \nabla \mathbf{q}=\mathbf{S}(\mathbf{q}),
\end{equation}
where $\textbf{q}$ is the vector of state variables, $S$ represents the vector 
of source terms, $\mathbf{F}$ are the conservative terms, and $B$ the 
non-conservative terms. In particular, for system \eqref{adult4} we have $B=\textbf{S}=0$.

The presented examples are solved with the scheme of third order and the Rusanov 
flux. For laminar flow and the Gresho vortex, periodic and transitive boundary 
conditions are applied, respectively. Parameters are chosen as $\kappa=1$ and 
$\alpha=1$ in Eq. \eqref{generic_a}. Computational examples run till the 
final time $T_{\mathrm{final}} = 0.5$ in 100 time steps on a $100\times100$ 
mesh. Let us now discuss the laminar channel flow. 

\subsubsection{Laminar channel flow}
The first example corresponds to the setting in Sec. \ref{sec:reduced_model}.
The computation domain is $\Omega=[0, 2\pi] \times [-\pi/2, 3\pi/2]$
and initial conditions are
\begin{equation}
\rho_1 = 0.5 + 0.25 \sin(x),
\end{equation}
$\rho$ and $\vv$ being prescribed by formulas \eqref{init}.

Figure \ref{fig1} shows the numerical solution to the system \eqref{rho4}, 
\eqref{u4}, \eqref{rhojfinal}.
Similar profiles of SHTC and GENERIC solutions at times $T_{\mathrm{final}}$ 
(Fig.~\ref{fig1} 
top-right) and $T_{\mathrm{final}}/2$ (Fig.~\ref{fig1} middle-bottom), 
respectively, indicate that advection within the reduced GENERIC model is (with 
the chosen material parameters $\kappa$ and $\alpha$) twice faster than in the 
reduced SHTC model.


\subsubsection{Gresho vortex}
\begin{figure}[ht]
  \begin{center}
    \begin{tabular}{ccc} 
      \includegraphics[trim = 50 20 50 
      10,clip,width=0.3\linewidth]{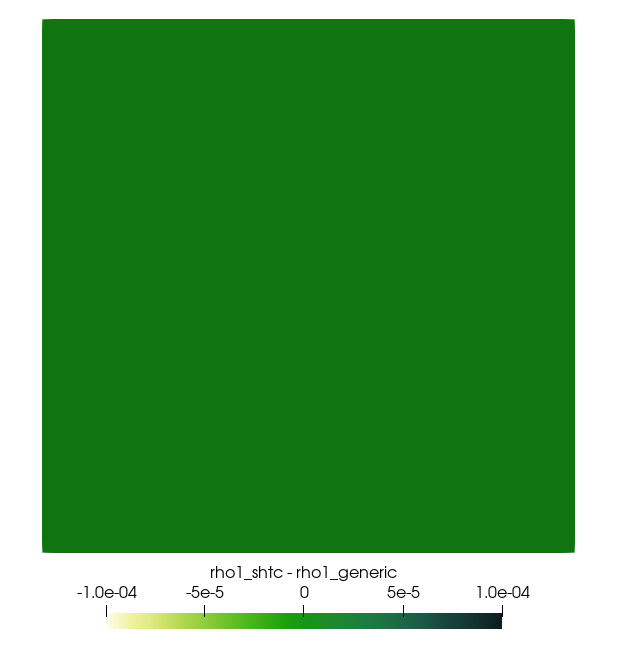} &
      \includegraphics[trim = 50 20 50
      10,clip,width=0.3\linewidth]{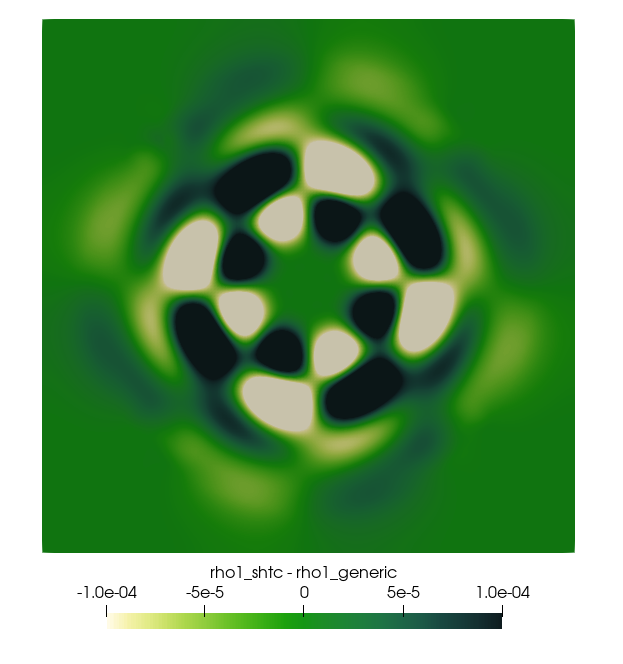} &
      \includegraphics[trim = 50 20 50 
      10,clip,width=0.3\linewidth]{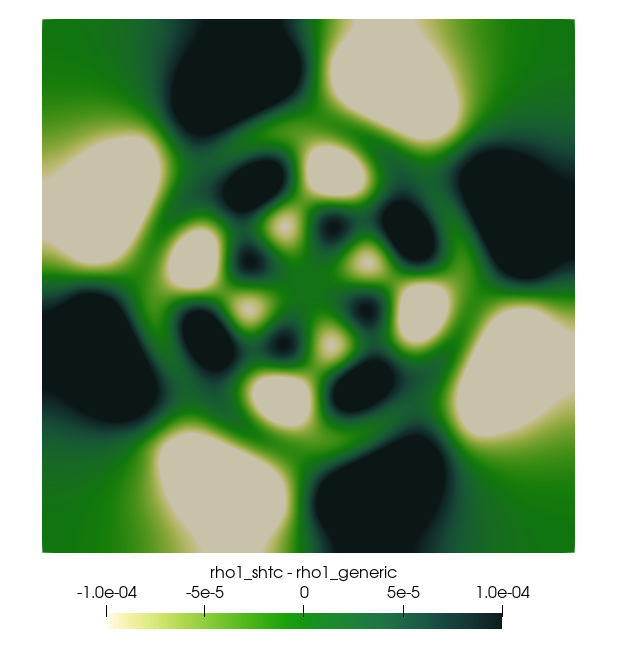} \\
      \includegraphics[trim = 50 20 50
      10,clip,width=0.3\linewidth]{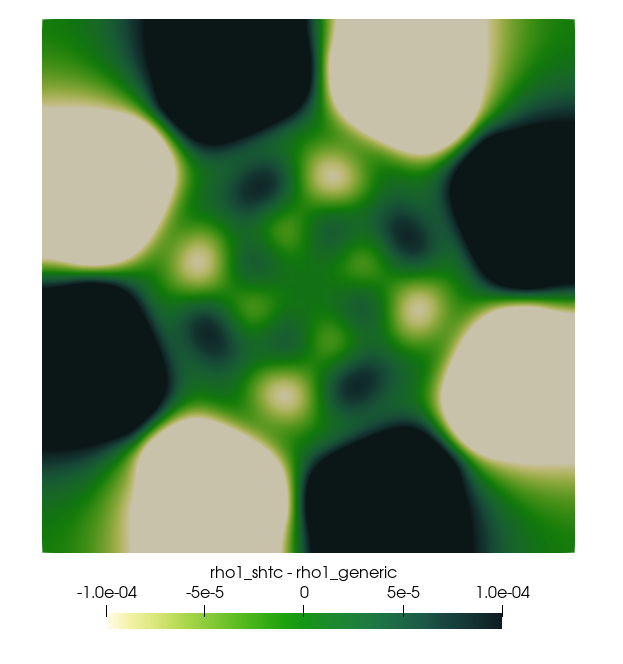} &
      \includegraphics[trim = 50 20 50
      10,clip,width=0.3\linewidth]{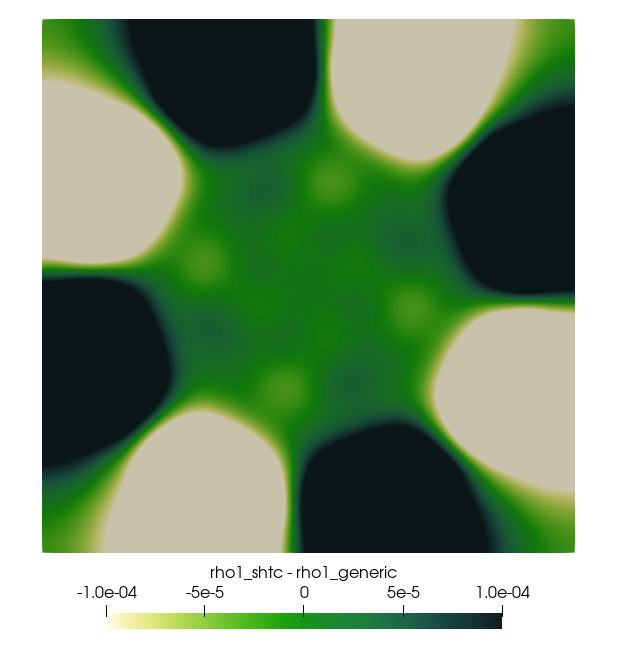} &
      \includegraphics[trim = 50 20 50
      10,clip,width=0.3\linewidth]{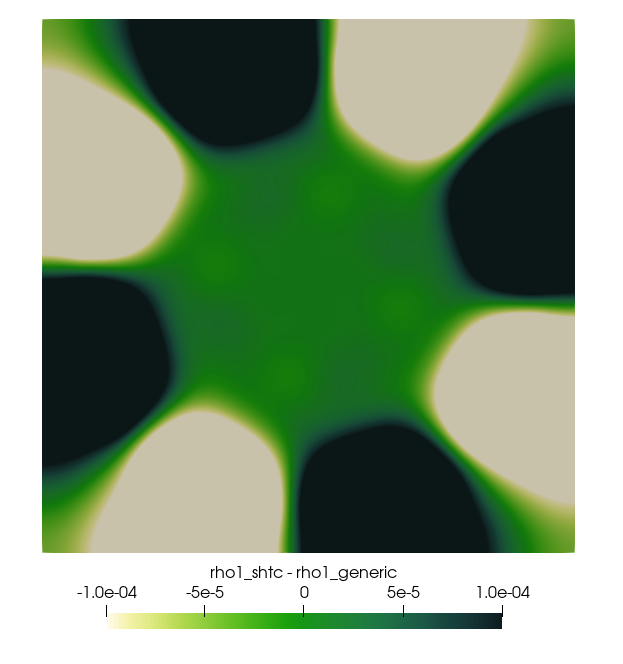}
    \end{tabular}
    \caption{Time evolution of the difference between $\rho_1$ in SHTC model 
      and the GENERIC model, $\rho_1^{SHTC} - \rho_1^{GENERIC}$ for initial 
      data of the Gersho vortex. We present six 
      times, in the reading order, $t = 
      0,1/5T_{\mathrm{final}},2/5T_{\mathrm{final}},
      3/5T_{\mathrm{final}},4/5T_{\mathrm{final}},T_{\mathrm{final}}$)} 
    \label{fig2}
  \end{center}
\end{figure}

The second computational example is based on the Gresho vortex for the Euler 
equation, where the centrifugal force is balanced by the pressure gradient, 
resulting in a stable time-independent vortex \cite{Gresho1990}. In our case, 
the mixture density, plays the role of the pressure, as can be in from Eq. ~\eqref{u4}. 

Let the computation domain be a square $\Omega=[0, 1] \times [0,1]$.
Initially we prescribe a Gresho body vortex outside of the origin,
\begin{align}
\rho &= \left\{\begin{array}{ll}
5 + \frac{25}{2} r^{2}, & r<0.2 \\
9 -4 \ln(0.2) + 4 \ln (r) -20 r+ \frac{25}{2} r^2, & 
0.2 \le r \le 0.4 \\ 
3 + 4 \ln (2), & 0.4 < r 
\end{array}\right. , \\
\rho_1 &= 2.5 , \\
\mathbf{v} &= \boldsymbol{e}_{\phi}\left\{\begin{array}{ll}
5 r, & r<0.2 \\
2-5 r, & 0.2 \le r \le 0.4 \\
0, & 0.4 < r
\end{array}\right. ,
\end{align}
where $r=\sqrt{(x-0.5)^{2}+(y-0.5)^{2}}$ and 
$\boldsymbol{e}_{\phi}=\left(-(y-0.5),x-0.5\right)/r$.

Figure \ref{fig2} shows the time evolution of the difference in the density of the first component 
$\rho_1^{SHTC} - \rho_1^{GENERIC}$.
We can see that the vorticity-dependent 
diffusion matrix within the reduced GENERIC model causes extra interaction 
between the vortex and $\rho_1$. 
Field $\rho_1$ now flows partly in the 
direction perpendicular to the (radial) gradient of chemical potential, that is  
also in the azimuthal direction. Those extra terms break the radial symmetry of 
the solution\footnote{More precisely, the symmetry is broken by the boundary 
conditions, but the extra terms highlight this symmetry breaking in the 
solution.}.
Comparison with experimental data is, however, out of scope of the present paper.

\section{Conclusion}
The main goal of this manuscript is to compare two recent frameworks of 
continuum thermodynamics, the SHTC equations and GENERIC, in the context of binary homogeneous mixtures. 
Using GENERIC, we can start from the Liouville equation and obtain equations for binary mixtures with additional 
terms that are not present in the SHTC framework, namely the terms containing the self-advection of the relative velocity. 

These terms can be analysed on simplified versions of the SHTC and GENERIC 
models, letting the relative velocity $\ww$ relax to a quasi-equilibrium value. 
Within GENERIC, we obtained a vorticity-dependent diffusion matrix.
Consequently, the density of the first constituent flows not only against the 
gradient of the chemical potential but also in the 
direction perpendicular to the gradient and the vorticity pseudovector.
This implies for instance, different advection rates in the two models or 
symmetry breaking of the solution in the presence of vorticity, see Figs. 
\ref{fig1} and \ref{fig2}.

Apart from the terms with self-advection of the relative velocity, the mixture models can have two entropies, the overall entropy, or no entropy at all.
The first option is observed for instance in cold plasma, where light electrons and heavy ions absorb electromagnetic energy with different efficiency.
Second possibility assumes that the components have the same temperature, and the third option is isothermal. 
The first option (two entropies) has uniquely determined Poisson bracket, that is derived from the Liouville equation. 
The second option (one entropy) is, however, much more difficult in the context of Hamiltonian mechanics. But we have shown that there is a family of 
 Hamiltonian binary mixture model with only the overall entropy 
that still contains the self-advection terms of the relative velocity while satisfying the Jacobi identity. The entropy difference $\Delta s$ 
is given by solutions to \eqref{partial}, for example by formulas \eqref{cond}, \eqref{cond2}. Finally, the isothermal case then again has a unique Poisson bracket. 

The structure of the SHTC model \eqref{eq.SHTC_full}, which does not contain 
the self-advection terms, is also Hamiltonian (considering only the reversible 
part) and furthermore admits constant $\Delta s$ (without violating Jacobi identity), unlike the GENERIC model.

In future, we would like to compare the results with 2D experimental results on 
diffusion in the presence of vorticity. Furthermore, it remains to generalise 
the analysis to heterogeneous mixtures, taking volume fractions among the state variables.



\section*{Acknowledgment}
M.P. and M.S. were supported by Charles University Research program No. 
UNCE/SCI/023.
M.P., V.K., and M.S. were supported by Czech Science Foundation, project no. 
20-22092S.
M.S. was supported by Grant Agency of Charles University, student project no. 
282120. PM acknowledges the financial support of the GRK 2297 MathCoRe, funded 
by the Deutsche Forschungsgemeinschaft, grant number 314838170.
The work of E.R. was supported by Mathematical Center in Akademgorodok, the agreement with Ministry of Science and High Education of the Russian Federation number 075-15-2019-1613.
I.P. is a member of the INdAM GNCS group and acknowledges the financial support received from  
the Italian Ministry of Education, University and Research (MIUR) in the frame of the Departments 
of Excellence  Initiative 2018--2022 attributed to DICAM of the University of Trento (grant L. 
232/2016) and in the frame of the 
PRIN 2017 project \textit{Innovative numerical methods for evolutionary partial differential 
equations and  applications}.


\renewcommand{\refname}{\spacedlowsmallcaps{References}} 

\end{document}